# DEBH: Detecting and Eliminating Black Holes in Mobile Ad Hoc Network

Ali Dorri
Department of Computer Engineering,
Mashhad Branch, Islamic Azad University, Mashhad, IRAN.
Alidorri.ce@gmail.com

Soroush Vaseghi
Department of Computer Science, Faculty of Engineering, University of Bojnord, Bojnord, Iran.
Sorouush@gmail.com

Omid Gharib
Department of Computer Engineering,
Mashhad Branch, Islamic Azad University, Mashhad, IRAN.
omidgharib1990@gmail.com

**Abstract**

Security in Mobile Ad hoc Network (MANET) is one of the key challenges due to its special features e.g. hop-by-hop communications, dynamic topology, and open network boundary that received tremendous attention by scholars. Traditional security methods are not applicable in MANET due to its special properties. In this paper, a novel approach called Detecting and Eliminating Black Holes (DEBH) is proposed that uses a data control packet and an additional Black hole Check (BCh) table for detecting and eliminating malicious nodes. Benefiting from trustable nodes, the processing overhead of the security method decreases by passing time. Ad hoc On-demand Distance Vector (AODV) routing protocol is used as the routing protocol in our design. After finding the freshest path using AODV, our design checks the safety of selected path. In case of detecting any malicious node, it is isolated from the entire network by broadcasting a packet that contains the ID of malicious nodes. Simulation results show that DEBH increases network throughput and decreases packet overhead and delay in comparison with other studied approaches. Moreover, DEBH is able to detect all active malicious nodes which generates fault routing information.

**Keywords**

Black hole Attack, Malicious Node, Routing Attack, Security Challenges.

## 1. Introduction

A set of self-configurable wireless nodes in a dynamic mode without any fix infrastructure or centralize management is termed as Mobile Ad hoc Network, usually known as MANET. MANET is popular in applications like emergency rescue, humanitarian aid and military due to its special characteristics including fast and easy implementation, hop-by-hop communications and mobile nodes [1]. Because of its special features, MANET faces with diverse types of challenges including routing [2], dividing mobile nodes into clusters [3] and providing security [4]. Among MANET's challenges, security is the most critical challenge due to features like open network boundary, dynamic topology, hop-by-hop communications and wireless media [5]. A comprehensive review on MANET's security challenges has been proposed in our previous work [6].

Ad hoc On-demand Distance Vector (AODV) is a reactive routing protocol, which is widely used in MANET [7]. AODV routing protocol is highly vulnerable against routing attacks, especially black hole. In black hole, malicious nodes inject fault routing information in order to persuade the source node to opt the path with malicious node as the best path. By receiving data packets, malicious node destroys them all and causes a Denial Of Service (DOS) attack [8].

In this paper, we propose a novel approach to detect and eliminate black hole attack in AODV-based MANET called Detecting and Eliminating Black Holes (DEBH). In DEBH a Black hole Check (BCh) table is kept by each node which assists nodes in detecting and eliminating malicious nodes. Beside ordinal control packet, DEBH uses a data control packet in order to detect malicious nodes in a path. DEBH detects all malicious nodes in network and eliminates them by low packet overhead and delay, By use of BCh and data control packet. Moreover, it increases network throughput by isolating all malicious nodes. DEBH detects and isolates all active malicious nodes with any position and order in network, just by first execution. By tacking advantages of trustable nodes, delay and packet overhead of DEBH is decreased by passing time. To show the advantages of DEBH, different scenarios are carried out in Opnet Modular 14 simulator. Simulation results show that DEBH's packet overhead and delay are lower than previous works.

The rest of the paper is organized as follows: Section 2 provides a brief review on AODV routing protocol and black hole attack. Section 3 discusses literature review. The DEBH is proposed in section 4. Section 5 provides simulation results and finally, section 6 concludes the paper.

**2. Study of AODV and Black hole**

In this section, we present a brief overview of AODV routing protocol and black hole attack.

2.1 Ad hoc On Demand Distance Vector (AODV) Routing Protocol

AODV is a routing protocol initiates the route discovery on-demand. Generally, on-demand routing protocols discover a route whenever it needed for packet transmission. Due to this feature, AODV categorized in reactive routing protocols [9]. In this protocol, every mobile node maintains a routing table and use it to find its Next_Hop_Node (NHN) to the destination. Each time a source node wants to send packets to a destination, it has to check its routing table. If the source has a fresh enough route to the destination, it will send packets through existing path. Otherwise, it has to find a route by using two control packets which are: Route Request (RREQ) and Route Reply (RREP). The source node initiates a route discovery process by broadcasting a RREQ packet to its neighbors. On receiving RREQ packet, Intermediate Nodes (INs) update their routing table for reverse path, then generate a RREP or rebroadcast the RREQ packet. A RREP packet is generated when either the IN is itself the destination or it has a fresh enough route to the destination. Otherwise, the IN increases RREQ's hop count and broadcast it again. The IN that generates RREP packet unicast it for the source node. All INs which receive RREP packet, update their routing tables and forward RREP toward the source node.

Receiving more than one RREP by the source node is strongly possible since all communications are hop-by-hop and RREQ packets are broadcast in network [9]. Among receiving packets the best path is the path with the highest sequence number which is defined as the freshest path [10]. Sequence number is increased by either an IN node which generates RREP or the source node that generates RREQ.

2.2 Black hole Attack

Black hole attack is a kind of Denial of Service (DOS) attack in which malicious node uses routing protocol's vulnerability and leads all data packets toward itself. Therefore, the way that each malicious node uses in order to break into the network is differ based on the network's routing protocol [11]. In an AODV based network, malicious node generates a RREP packet with a high sequence number in response to RREQ packets. Consequently, the path with malicious node is opted as the freshest path by the source node.

Regarding the number of malicious nodes participating in network, black hole can be studied in three types which are: Single, Cooperative and Distributed black hole. In single black hole, as showed in Figure 1.a, there is just one malicious node in network; while, in cooperative black hole, as showed in Figure 1.b there are more than one malicious nodes that cooperatively work with each other in order to cover their tracks. In this type of attack malicious nodes are in the wireless range of each other. In addition to these types, we defined a new type of black hole, which is distributed black hole attack. In this type of attack, malicious nodes are distributed and can be in different locations in network. All malicious nodes are aware of other malicious nodes' position and ID and work cooperatively to cover their tracks.

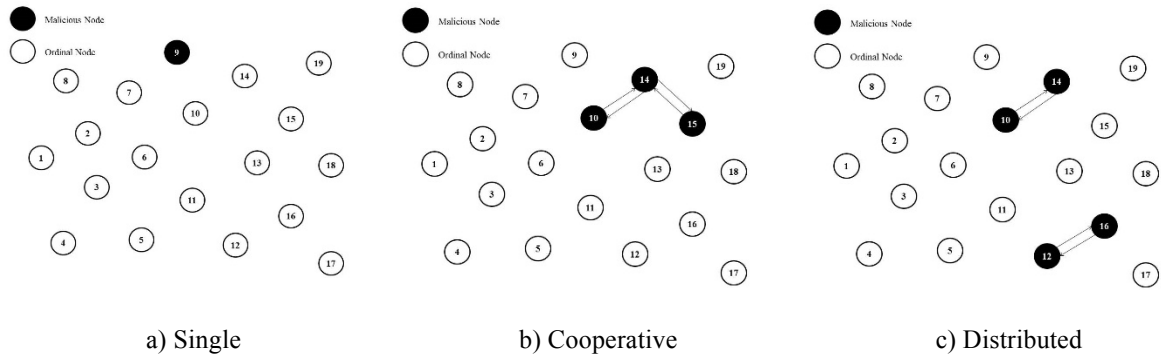

a) Single             b) Cooperative            c) Distributed

Figure 1. Different types of Black hole attack.

The number of malicious nodes in each location can be different. Figure 1.c. is an example of distributed black hole attack.

## 3. Related Works

In literature many black hole detection and/or elimination schemes have been presented. Authors in [12] presented an approach based on multi-path packet sending. In this approach, packets are divided to some sub-packets and then each sub-packet is sent through a random path to the destination. As a result, malicious node can access limited number of packets sent through its path. Simulation results proof that the proposed approach is effective in increasing packet delivery; however, dividing packets and sending them through multipath routes increases network overhead and processing time in the destination. Since the destination has to cache all packets until all of them are received.

Authors in [13] presented a new approach in order to detect and eliminate black hole attack without using any extra packet or additional packet headers. Referring to this approach, each time a source node receives RREP packets, it generates a new RREQ and put the best received sequence number as the new RREQ packet's sequence number and unicast it through the route which the RREP packet was received. By receiving RREQ packet, malicious node generates a RREP packet with higher sequence number than received one. The malicious node sends fault RREP packet to the source node. As the malicious node sends a sequence number higher than its previous sequence number, the source node mark RREP generator as malicious. This approach can detect malicious node without using any additional packet, however, it can detect just the RREP generator and is not able to detect cooperative malicious nodes.

Authors in [14] presented a new approach based on confirming the best path using second path. In this approach, whenever a source node receives RREP packets, it sends a confirmation packet through the second best path to the destination and asks the destination whether it has a route to the RREP generator or to the Next_Hop_Node of RREP generator or not. If the destination has no route to these nodes, both RREP generator and its Next_Hop_Node are marked as malicious nodes. Using this approach, the source node can detect cooperative malicious nodes. However, this method cannot detect more than two malicious nodes.

Authors in [15] presented an approach based on promiscuous mode in INs. Using this approach, each node monitors its neighbors and calculates a threshold to detect malicious nodes. The threshold is a ratio between received packets and forwarded packets. This approach can detect one malicious node; however, it is unable to detect cooperative malicious nodes as they send data packets to each other to bypass the security method.

In our previous work [16] we proposed a security method based on checking the NHN and Previous_Hop_Node of each RREP generator. Each source node checks both nodes after and before RREP generator using a control packet and a Data Routing Information (DRI) table. This approach increases throughput and decreases delay and packet overhead. However, it is unable to detect distributed malicious nodes.

In [17] authors proposed a trust based AODV protocol named TAODV. In this method, each node maintains a trust table indicating three trust levels, namely, unreliable, reliable, and most reliable. The unreliable nodes are new nodes just join to the network or nodes which have no history. Reliable nodes are nodes that communicated some packets with the current node, and finally most reliable nodes are those who transmitted many packets with the neighbor. The trust rating is increased based on the number of packets sent and received by the neighbor. Simulation results shown that the proposed method has higher packet delivery compared to normal AODV. However, the paper suffers from lack of simulation results as the results are compared only with normal AODV.

In [18] authors proposed a security method which uses a time metric on the source side node to protect the network against malicious black hole nodes. In this method, the destination sends RREP for all RREQ packets it receives. The source node collects RREP packets based on time and number of hops and sends data packet either through the first path or divides it and sends it through the first five paths. If data is not received in a path to the destination, then the source send data using another path.

In this section a literature review of existing detection and/or elimination approaches for black hole attack has been discussed. Generally, it can be inferred that using promiscuous-based approaches in order to detect malicious nodes is not affective, as the malicious nodes send data packets between each other in order to bypass the security mechanisms. In addition, lots of existing approaches can detect only single or cooperative attack with two malicious nodes. These limitations decrease MANET flexibility and make it vulnerable against black hole attack. In the other hand, these approaches increase delay and packet overhead and suffer from low network throughput.

## 4. Detecting and Eliminating Black Holes (DEBH)

In order to overcome limitations of existing detection and/or elimination approaches, we propose a new approach called Detecting and Eliminating Black Holes (DEBH) to detect all malicious nodes in any order and any position in network. The DEBH uses an additional data control packet in order to detect malicious nodes. We previously proposed this data control packet in [19]. In our previous work, this data control packet was used once by the source node for checking the safety of selected path. The structure of the data control packet is the same with the previously designed one; however, in DEBH the data packet is used for checking path in all steps and for all nodes. In addition, each node keeps a Black hole Check (BCh) table to determine trustable nodes. Lots of detection approaches commence security algorithm from the RREP generator and all nodes between the source and the RREP generator are assumed to be safe. However, because malicious nodes are cooperative, it is possible that the last malicious node generate RREP in order to cover its cooperatives.

The DEBH uses two different queues for checking the nodes in paths, which are; "Black hole" queue and "RREP generator" queue. "Black hole" queue contains the ID of nodes which are suspected to be malicious and "RREP generator" queue contains the ID of nodes which has generated the freshest RREP in response to RREQs. The DEBH approach consists of the following four phases: 1) Finding the freshest path. 2) Path security analysis. 3) BCh update. 4) Eliminating malicious nodes.

**Phase 1) Finding the freshest path**

This phase is based on AODV routing protocol which was discussed in section 2. The only difference is that in the DEBH when an Intermediate Node (IN) wants to generate RREP packet, it has to send its Next_Hop_Node (NHN) and its BCh entry for NHN within the RREP packet to the source node. The source node uses this information to find malicious nodes.

**Phase 2) Path security analysis**

After choosing the freshest path and before sending data packets, the source node has to ensure that the selected path is safe. Figure 2 shows this phase in summary. For doing this, the source node uses a data control packet and IN's BCh table. The structure of data control packet is shown in Figure 3 [19].

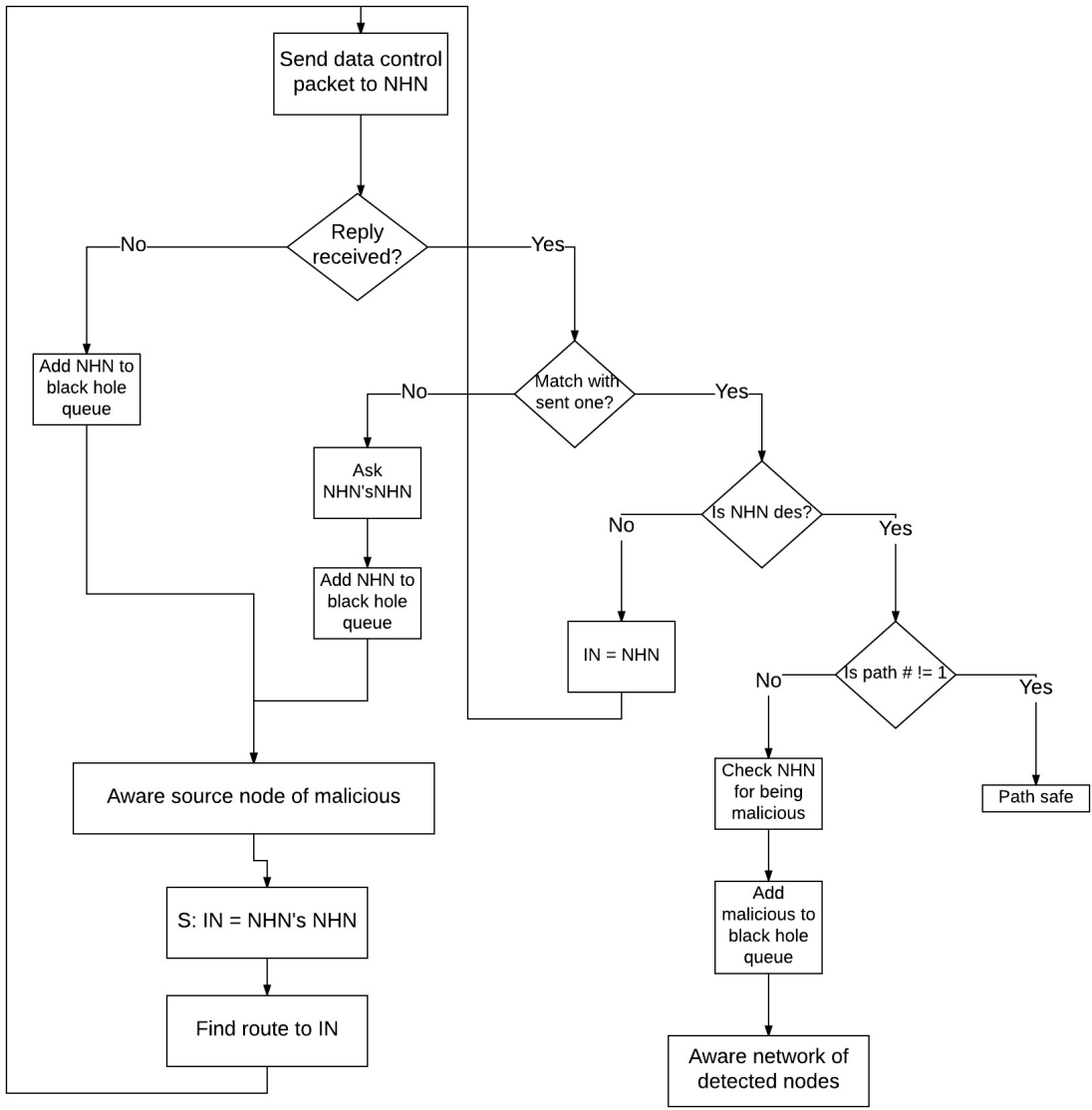

Figure 2. A summary of detecting malicious nodes.

Figure 3. Data control packet [19].

This data control packet consists of the following parameters:

**Node_ID:** This parameter refers to the ID of data control packet's generator. In DEBH each node has to discard the received data control packet and generate a new one based on its own properties.

**NHN:** This parameter refers to data control packet generator's NHN in the path toward the destination.

**Random_Number:** After establishing the security analysis phase, the source node generates a random number and put it in the data control packet. INs have to fetch this number and put it into their own data control packet. Therefore, this number remains the same for all generated data control packets; however, for checking new path, a new random number is generated. Since this control packet is a kind of data packet, malicious nodes can't relay it to ordinal nodes. The proposed BCh table is shown in Figure 4.

| Node ID | Trustable |
|---------|-----------|
| 2 | 0/1 |
| 5 | 0/1 |
| 1 | 0/1 |

Figure 4. Black hole Check (BCh) table

Each IN updates this table using the proposed data control packet in the security analysis phase. We will discuss the updating process later in this section. In this table "Node ID" refers to the node's identification and "Trustable" refers to whether the node is trustable for host node or not. Definition 1 defines the trustable nodes.

*Definition 1: Node B is trustable for node A if and only if node A sends data control packet and receives reply from node B.*

The process of checking path is illustrated in algorithm 1.

**Algorithm 1:** The process of path security analysis

```
1:Path_Number=1
2:Source: Generate a random number
3:Add RREP generator to RREP generator's queue
4:Source: Set source node as IN
5:IN: Generate data control packet and send it for NHN
6:IN: Wait for reply
7:IN: If reply is received
8:      {
9:         If received random number is the same with sent number
10:            {
11:               Update BCh table
12:               If NHN is destination
13:                  {
14:                     If Path_Number != 1
15:                        {
16:                           Ask NHN's BCh entry
17:                           Check Malicious
18:                           If is malicious
19:                              {
20:                                 Add it to Black hole queue
21:                                 Go to 52
22:                              }
```

```
23:                            }
24:                  Path is safe
25:                  End:
26:               }
27:            Set NHN as IN
28:            Go to Line 5
29:            }
30:      Else
31:            {
32:            Ask NHN's NHN
33:            Add NHN to Black hole Queue
34:            Go to Line 42
35:            }
36:   }
37:Else
38:      {
39:      Add to Black hole Queue
40:      Go to Line 42
41:      }
42:IN: Aware Source node of NHN's ID
43:Source: Set NHN's NHN as IN
44:If IN is RREP generator
45:      {
46:      Set RREP generator's NHN as IN
47:      Go to Line 49
48:      }
49:Source: Find a route to IN
50:Path_Number ++
51:Go to 2
52:Source: Aware network of detected malicious nodes
```

By opting the best RREP, the source node adds the RREP generator to "RREP generator" queue. Initially, the source node generates a random number and sets it in the data control packet. Then the generated data control packet is sent for the source node's NHN (lines 1-5). By receiving this packet, each IN has to extract the random number and generates a new data control packet with its own properties, then sends it to its previous node and NHN. The packet which is sent for the previous node, is considered as the reply for data control packet. This process is continued until one of the situations below happen:

1) NHN is destination: In this case the path is safe and the destination sends an ACK to the source node (lines 12,24).

2) Received random number is not the same with the sent one: In this case, the IN has to ask its NHN's NHN by using an ordinal control packet and sends it to the source node (lines 30-35).

3) After a period of time, reply is not received: This is similar to situation 2 (lines 37-41).

If one of the situations 2 or 3 happen, the source node become aware by IN, then the IN's NHN is added to "Black hole" queue by the source node as they are suspected to be malicious. Then the source node has to find a new path to the suspected node's NHN and ask it for its own NHN and BCh entries for its own NHN and previous node (IN) (lines 43-50). For this, the source node starts from phase one of DEBH and uses data control packet to check the safety of all nodes in the path through its new destination. A "Path_Number" is defined in order to examine whether it is the first checking path or not. In case that it is greater than '1' it means if packet reaches to the destination, the truth of claims of previous nodes have to check by using BCh table (lines 14-23). The source node checks the malicious nodes by using Definition 2.

*Definition 2: Node A is malicious if its BCh entry for node B has been set as '1' and node B's entry for node A has been set as '0'.*

In addition to nodes which don't send reply for data packet, RREP generators are also suspension to be malicious; therefore, the source node adds them in a separate queue which is "RREP generator" to later evaluate their truth. If each node in this queue claimed that it has communicated with malicious nodes, then this node is marked as malicious node. For connecting to new NHN, the source node and all INs use data control packet; therefore, if malicious node send fault information, it is detected by INs.

Sending data packets hop-by-hop, increases delay and packet overhead. In order to overcome this challenge if an IN has trust to its NHN, there is no need to send data packet and it can use ordinal control packet. Based on Definition 1, if node A is trustable for node B, node B is trustable for node A.

The discussed process continues until one of the RREP generators (sent RREP in response to line 49) send ACK. In this situation the source node trusts RREP generator and checks nodes in both queues. The detail of checking queues is presented in phase 4.

By following the described process all malicious nodes are detected. For more clear illustration, an example is given. Consider the network in Figure 1.b. Node '1' is the source node. Table 1 propose the rules (in Algorithm 1) and parameters in this network.

Table 1. Steps of proposed approach for cooperative network in Figure 1.b.

| Node's ID | NHN's ID | Path_Number | RREP generator Queue | Black hole Queue | Rules Followed |
|---|---|---|---|---|---|
| 1 | 2 | 1 | 15 | ------ | 1-12,27,28 |
| 2 | 10 | 1 | 15 | ------ | 5-7,37-44,49-51 |
| 1 | 3 | 2 | 15,14 | 10 | 5-11,27-28 |
| 3 | 6 | 2 | 15,14 | 10 | 5-11,27-28 |
| 6 | 11 | 2 | 15,14 | 10 | 5-11,27-28 |
| 11 | 13 | 2 | 15,14 | 10 | 5-11,27-28 |
| 13 | 14 | 2 | 15,14 | 10 | 5-7,37-51 |
| 1 | 3 | 3 | 15,14,3 | 10,14 | Trustable, 12-22 |

As another example consider network in Figure 1.c. Table 2 propose the rules and parameters in this network.

Table 2. Steps of proposed approach for distributed network in Figure 1.c.

| Node's ID | NHN's ID | Path_Number | RREP generator Queue | Black hole Queue | Rules Followed |
|---|---|---|---|---|---|
| 1 | 2 | 1 | 14 | ------ | 1-12,27,28 |
| 2 | 10 | 1 | 14 | ------ | 5-7,37-51 |
| 1 | 3 | 2 | 14,16 | 10 | 5-11,27-28 |
| 3 | 11 | 2 | 14,16 | 10 | 5-11,27-28 |
| 11 | 12 | 2 | 14,16 | 10 | 5-11,27-28 |
| 1 | 8 | 3 | 14,16,8 | 10,12 | 5-22 |

**Phase 3)** Black hole Check (BCh) Table

In DEBH, each node keeps a low size table called Black hole Check (BCh) for all neighbors. Due to MANET dynamic topology each IN's neighbor may change over time, however each IN keep record of all its previous neighbors. By receiving data control packet's reply both IN and its NHN update their BCh tables and set "trustable" column as '1'. This means these two nodes have trust to each other. For decreasing packet overhead and delay, trustable nodes have to use ordinal control packets for checking the path. This ordinal packet just sent toward the destination and sending reply is not needed. By tacking advantages of the BCh tabl packet overhead and delay decreases and it even may reach to zero when RREP generator is trustable for the source node. Another advantages of BCh table is its low size in compare to other tables like DRI table [16].

**Phase 4) Eliminating Malicious Nodes**

At the end of phase 2 all malicious nodes are marked. DEBH uses another queue which is "RREP Generator" queue. RREP generators are suspected to be malicious and at the end of the security algorithm, they are checked by the source node. This queue is First In First Out (FIFO) and if the first node has been marked as malicious, all other nodes are marked as malicious. The reason is that, they claimed to have communication with marked node. By checking nodes in "RREP Generator" queue, one of the following situations will happen:

    1) RREP generator's NHN is trustable and has no way to RREP generator: In this case, RREP generator is

marked as malicious node.

2) RREP generator's NHN is in black hole queue: RREP generator is marked as black hole since it claims to have communications with a malicious node.

3) RREP generator's NHN is trustable and has a way to RREP generator: RREP generator is marked as trustable and path is safe.

Finally, the source node puts all malicious nodes ID in a packet and broadcast it to network. By receiving this packet each IN puts "Trustable" column as "NULL" for enounced nodes, then the packet is rebroadcasted. By now, if an IN receives a RREP packet from malicious nodes, the packet will be discarded.

## 5. Evaluation

In this section experimental setup, performance metrics, simulation results and analyses are discussed.

### 5.1 Experimental Setup

We simulate our approach and other works in Opnet 14 simulator to evaluate the performance of our method. Three works were implemented which are: 1) DEBH, 2) A watch dog mechanism proposed in [20] which is called "watchdog" in the rest of the paper, 3) EDRI approach proposed in [19], which is called EDRI in the rest of the paper. The goal of this paper is to provide an effective approach for detecting malicious nodes with any order and in any position in the network. Therefore, each approach has been implemented in different scenarios which are as follow: 1) Single black hole, 2) Cooperative black hole with two nodes, 3) Cooperative black hole with three nodes, 4) Cooperative black hole with five nodes, 5) Cooperative black hole with seven nodes, 6) Cooperative black hole with nine nodes, 7) Distributed attack with two malicious nodes in two different path.

Initially nodes are randomly distributed in an area of 1000 m * 1000 m. We use random waypoint model as mobility model and TCP traffic source. We use packet size of 512 bytes/packet. Since time required for rerouting has no effect on our approach's results, this time is passed up in our simulations. This time is important just in evaluation of routing protocols. Table 3 provides information on simulation parameters.

### 5.2. Performance Metrics

We use following metrics to evaluate the DEBH and other works:

**Packet overhead:** Due to wireless media, decreasing the number of control packets which are transmitted in network is highly demanded. We use the number of RREQ packets generated by the source node to evaluate this metric as the RREQ packets are broadcast to network.

Table 3. Simulation Parameters.

| Parameter | Value |
|---|---|
| **Simulation duration** | 600 sec |
| **Simulation area** | 1000*1000 |
| **Number of mobile nodes** | 30 |
| **Transmission range** | 200m |
| **Movement model** | Random waypoint |
| **Maximum speed** | 2-20 m/sec |
| **Traffic type** | TCP |
| **Packet rate** | 2 packets/sec |
| **Data payload** | 512 byte/packet |
| **Number of malicious nodes** | 2/3/5/7/9,4 |
| **Host paused time** | 15 sec |

**Delay:** Time taken by a security mechanism is important due to MANET dynamic topology. In our study, delay refers to time when the source node starts to find a route to the destination, until it found a secure path without any malicious nodes.

**Number of detected malicious nodes:** Since malicious nodes are cooperative, they may use some mechanisms in order to bypass security algorithms. Forwarding data packets between each other or generating RREP packet by the last node in the path, are some of covering approaches which are used by malicious nodes. This metric refers to each approach's accuracy. An accurate approach is an approach which detects all malicious nodes in a single run.

**Throughput:** This metric refers to the number of delivered packets in compare with sent packets in different number of connections.

### 5.3. Simulation Results and Analysis

Simulation results are presented in this section. Figure 5 proposed the simulation results for packet overhead in different situations.

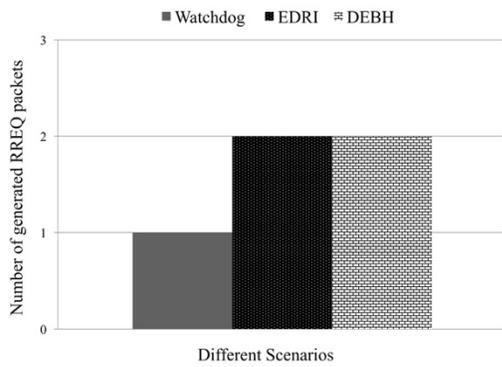
a) Single balck hole

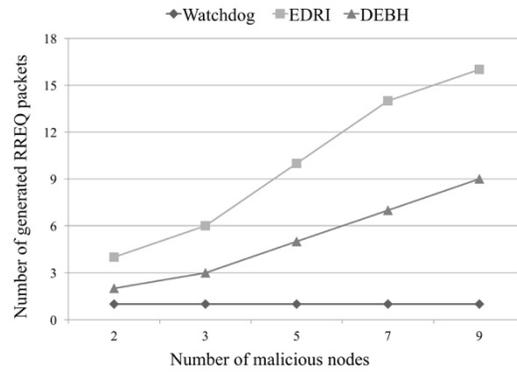
b) Cooperative black hole

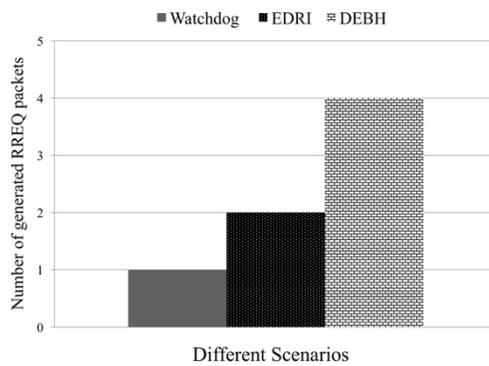
c) Distributed black hole

Figure 5. Simulation results for packet overhead

In these diagrams vertical axis refers to the number of RREQ packets which are generated by each security approach. As shown in Figure 5.a. watchdog approach's packet overhead is lower than both other approaches. The reason is that in watchdog mechanism, there is no need for nodes to generate any additional packet to detect malicious nodes. In other two approaches the source node has to generate a RREQ packet for connecting to the RREP generator's NHN. In cooperative attack, as shown in Figure 5.b, DEBH packet overhead is far lower than EDRI and the difference become greater by growth in the number of malicious nodes. Watchdog approach generates just one RREQ packet, however, it is not able to detect cooperative malicious nodes. Regarding distributed malicious nodes, as shown in Figure 5.c, DEBH generates far more packets in compare with other approaches; however, it's the only approach which can detect malicious nodes.

Simulation results to evaluate delay are shown in Figure 6. In these diagrams vertical axis refers to delay which is measured in second. In single black hole, as presented in Figure 6.a, watchdog mechanism detects malicious node far sooner than other two approaches. Since DEBH needs to send data packets hop-by-hop, delay of this approach is higher than others. As for cooperative attack, as presented in Figure 6.b, EDRI approach increases delay and DEBH detects all malicious nodes with far lower delay. Watchdog approach is not able detect malicious nodes. Regarding distributed black hole, as shown in Figure 6.c, DEBH has the highest delay; however, other approaches are not able to detect malicious nodes. This delay is incurred only once. The first node in network which detects malicious nodes informs other nodes. Additionally, using the trustable nodes decreases this delay for further communications significantly and the delay might even reach to zero.

Regarding the number of malicious nodes detected in each run, simulation results are given in Table 4. Watchdog mechanism is useful just for single black hole and cannot detect other types of attack. As for EDRI approach, it can detect malicious nodes one by one, since malicious nodes are cooperative and the last malicious nodes in each path generates RREP packet. The EDRI approach is not able to detect distributed black hole nodes. DEBH can detect all types of attack, as shown in Table 4.

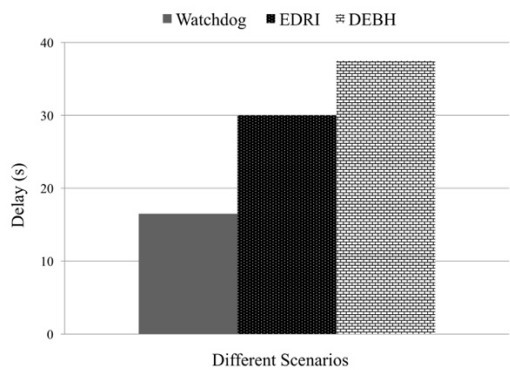
a)  Single Black hole

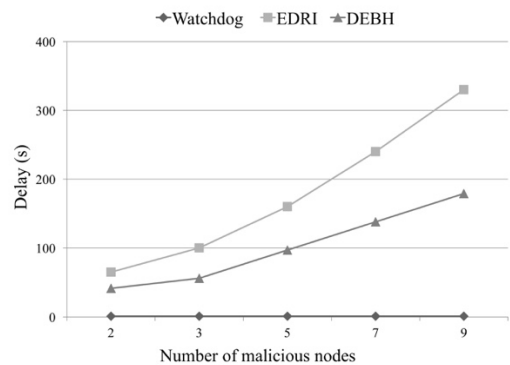
b)  Cooperative Black hole

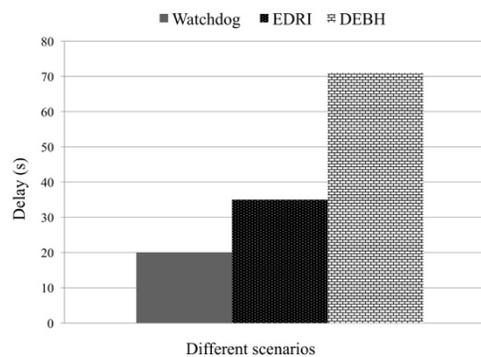
c)  Distributed Black hole
Figure 6. Simulation Results for Delay

Table 4: Evaluation of the number of detected malicious nodes.

| | Single Black hole | Cooperative Black hole | | | | | Distributed Black hole |
|---|---|---|---|---|---|---|---|
| | | 2 Malicious | 3 Malicious | 5 Malicious | 7 Malicious | 9 Malicious | |
| Watchdog | 1 | 0 | 0 | 0 | 0 | 0 | 0 |
| EDRI | 1 | 1 | 1 | 1 | 1 | 1 | 0 |
| DEBH | 1 | 2 | 3 | 5 | 7 | 9 | 4 |

Simulation results to evaluate the network throughput are shown in Figure 7. Since network throughput for cooperative attack with different number of malicious nodes is different, each of them are proposed separately. For more precise evaluation, different number of connections are used in network varied from 5 to 30 connections. In our study each malicious node in network sends RREP for just one node and it sends another RREP packet when all data packets from the previous node are received. In our study each node sends 10 packets during each connection.

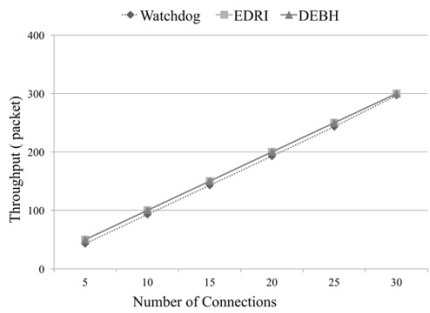

a) Single black hole

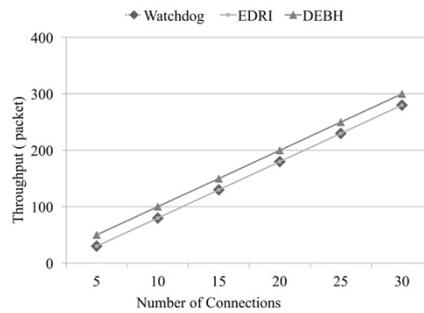

b) Cooperative with 2 malicious nodes

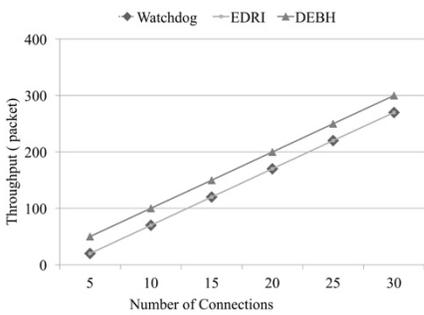

c) Cooperative with 3 malicious nodes

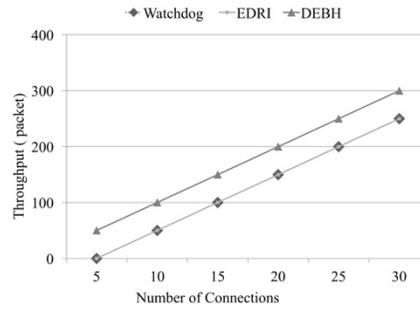

d) Cooperative with 5 malicious nodes

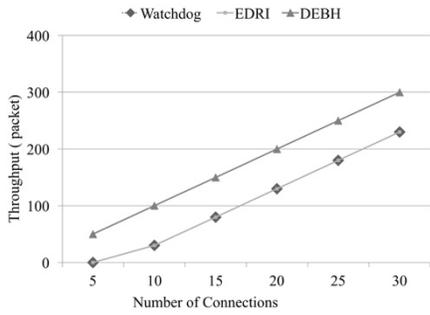
e)  Cooperative with 7 malicious nodes

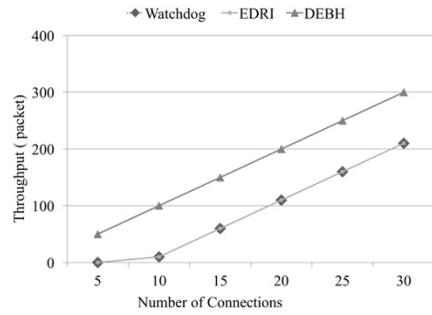
f)  Cooperative with 9 malicious nodes

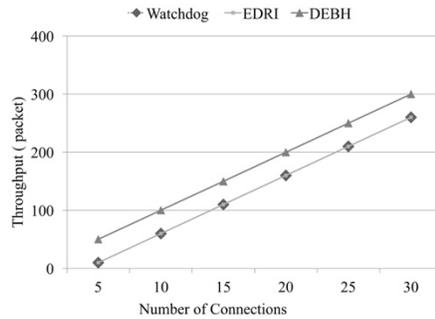
g)  Distributed black hole

Figure 7. Evaluation of network throughput

In single black hole attack as shown in Figure 7, DEBH and EDRI detect malicious nodes and deliver all packets to the destination, while watchdog needs a time to detect malicious node based on the number of dropped packets. In cooperative attacks as shown in Figures 7.b-7.f, each malicious node drops all received packets and other packets reach to the destination. Therefore, by increasing the number of malicious nodes in network, network throughput decreases significantly. Network throughput reaches to approximately 200 packets in the scenario with nine malicious nodes, while it is about 300 packets in scenario with two malicious nodes. Both watchdog and EDRI have the same throughput, since in both malicious nodes drop received packets. In EDRI approach at the end of the first run of algorithm and by detecting just one of the existing malicious nodes the source node starts sending data packets. Moreover, all connections have been started at the same time; therefore, in network with two malicious nodes each malicious node is detected by one source node in the same time. As for distributed black hole, since both EDRI and watchdog are not able to detect malicious nodes, the throughput of both approaches are the same, however, DEBH detects all nodes and deliver all data packets to the destination.

Referring to discussed results, it is clear that the DEBH is far more effective in cooperative and distributed attacks; however, it is not an effective choice for single black hole.

**Conclusion**

Mobile Ad hoc Network (MANET) is a kind of ad hoc network, which has mobile wireless nodes. In this network all nodes are free to move in network. Dynamic topology and hop-by-hop communications of MANET made security in routing protocols highly challengeable. In black hole attack, which is a kind of Denial of Service (DOS) attack, each malicious node uses the routing protocol's vulnerability and leads all data packets toward itself, then drops them all. Based on the number and position of malicious nodes in network, black hole attack is studied in three types which are: single, cooperative and distributed black hole. In this paper we proposed a novel approach called Detecting and Eliminating Black Holes (DEBH) which uses a data control packet and a Black hole Check (BCh) table for detecting and eliminating malicious nodes. By using data control packet BCh table is updated during the processing time of security mechanism and the number of trustable nodes increases dramatically. By increasing the number of trustable nodes, delay and packet overhead decreases significantly. Simulation results proof that our approach is able to detect all types of black hole, however in case of single black hole, packet overhead and delay caused by our approach is far greater than other studied approach.